\documentclass[12pt]{article}
\normalbaselines
\catcode `\@=11
\@addtoreset{equation}{section}

\def\section{\@startsection {section}{1}{\z@}{-3.5ex plus -1ex minus
     -.2ex}{2.3ex plus .2ex}{\normalsize\bf}}
\def\subsection{\@startsection{subsection}{2}{\z@}{-3.25ex plus -1ex minus
 -.2ex}{1.5ex plus .2ex}{\normalsize\bf}}
\def\thebibliography#1{\section*{References\markboth
  {REFERENCES}{REFERENCES}}\list
  {[\arabic{enumi}]}{\settowidth\labelwidth{[#1]}\leftmargin\labelwidth
  \advance\leftmargin\labelsep
  \usecounter{enumi}}
  \def\newblock{\hskip .11em plus .33em minus -.07em}
  \sloppy
  \sfcode`\.=1000\relax}
 
\catcode `\@=12
\def\be{\begin{equation}}
\def\ee{\end{equation}}
\def\bea{\begin{eqnarray}}
\def\eea{\end{eqnarray}}

\def\al{\alpha}
\def\bt{\beta}

\def\Ga{\Gamma}
\def\de{\delta}

\def\la{\lambda}

\def\si{\sigma}

\def\pr{\prime}
\def\pt{\partial}
\def\na{\nabla}

\def\tr{\hbox{\tt tr}}
\def\det{\hbox{\tt det}}
 
\begin{document}

\vspace*{2.5cm}
\noindent
{ \bf NONLINEAR LAGRANGIANS OF THE RICCI TYPE
\footnote{This work is supported by Polish KBN and partially 
by Mexican grant - proyecto de CONACyT \#27670 E
 (gr-qc/9906043)}}\vspace{1.3cm}\\
\noindent
\hspace*{1in}
%%%%%%%%%%%%%%%%%%%%%%%%
%   Use this for a multi-author contribution
%%%%%%%%%%%%%%%%%%%%%%%%
%\begin{minipage}{13cm}
%Arthur Author$^{1}$  and John Co-Author$^{2}$ \vspace{0.3cm}\\
% $^{1}$ Department of Mathematics, Bialowieza University, Poland \\
%      \makebox[3mm]{ }E-mail:  xx@yyyy.zzzz.ww
% \vspace{0.1cm} \\
% $^{2}$ Academy of Natural Sciences,\\
%\makebox[3mm]{ }USE - 99999 Asymptotic Region, United States of Europe\\
%\makebox[3mm]{ }E-mail:  aa@bbbb.ccc.dd
%\end{minipage}
%
%%%%%%%%%%%%%%%%%%%%%%%%
%   Use this for a single-author contribution
%%%%%%%%%%%%%%%%%%%%%%%%
\begin{minipage}{13cm}
Andrzej Borowiec\vspace{0.3cm}\\
Institute of Theoretical Physics, Wroc{\l}aw University, Poland \\
E-mail: borow@ift.uni.wroc.pl
\end{minipage}

\vspace*{0.5cm}

\begin{abstract}
\noindent
The Euler-Lagrange equations for some class of gravitational actions
are calculated by means of Palatini principle.
Polynomial structures with Einstein metrics appear among extremals
of this variational problem.
\end{abstract}

\section{\hspace{-4mm}.\hspace{2mm}
INTRODUCTION}

A {\it polynomial structure} on an $n$-dimensional differentiable manifold 
$M$ is given by type $(1,1)$ tensor field $S\equiv S^\mu_\nu$ of constant 
rank $r$ ($1\leq r\leq n$), which satisfies polynomial equation 
$\pi (S)=0$ for some polynomial $\pi (t)$ of real coefficients.
Almost-complex and almost-product structures are among
the best known examples and the most fundamental structures
of this kind \cite{Besse}. It has been recently shown that both 
these structures appear in a natural way from the first-order
(Palatini) variational principle applied to general class of
non-linear Lagrangians depending on the Ricci squared invariant
constructed out of a metric and a symmetric connection \cite{BFFV3}.
Moreover, Einstein equations of motion and Komar energy-momentum
complex are {\it universal} for this class of Lagrangians \cite{BFFV2}. 
The non-linear gravitational Lagrangians which still generate Einstein 
equations are particularly important since, at the classical level, they 
are equivalent to General Relativity. However, their quantum contents
and  divergences could be slightly improved. 

In the present  note, we are going to extend above
results showing that more general Ricci type Lagrangians
lead to more general polynomial structures and that the universality 
property remains still valid; both for the equations as for 
the energy-momentum. The techniques used here for analysis of the 
Euler-Lagrange equations are similar to the ones applied in 
\cite{BFFV2,FFV,FFV3} (c.f. \cite{BF} for summary). 
A different approach that missed polynomial relations has 
been recently proposed in \cite{TU}.

\subsection{\hspace{-5mm}.\hspace{2mm}Preliminaries and Notation}

Einstein metrics are extremals of the Einstein-Hilbert purely
metric variational problem.
It is known that the non-linear Einstein-Hilbert type Lagrangians 
$f(R)\sqrt g$, where $f$ is a function of one real variable and $R$
is a scalar curvature of a metric $g$
\footnote{One simply writes $\sqrt g$ for $\sqrt{|\det g|}$.},
lead to fourth order equations for $g$ which are not equivalent to 
Einstein equations unless $f(R)=R-c$ (linear case), or to appearance of 
additional matter fields. %\cite{}
It is also known that the linear "first order" 
Lagrangian $r\sqrt g$, where $r=r(g,\Ga)=g^{\al\bt}r_{\al\bt}(\Ga)$ is 
a scalar concomitant of the metric $g$ and  linear (symmetric) 
connection $\Ga$, \footnote{
Now, the scalar
$r(g,\Ga)=g^{\al\bt}r_{\al\bt}(\Ga)$ is not longer the scalar curvature, 
since $\Ga$ is not longer the Levi-Civita connection of $g$.} leads 
to separate  equations for $g$ and $\Ga$ which turn out to be equivalent 
to the Einstein equations for $g$ (so-called Palatini principle,
c.f. \cite{FFR,Lim,HB,Buch,HK}).

In the sequel we shall use lower case letters $r^\al_{\bt\mu\nu}$ and
$r_{\bt\nu}=r^\al_{\bt\al\nu}$ to denote the Riemann and Ricci tensor
of an arbitrary (symmetric) connection $\Ga$ 
\bea
r^\al_{\bt\mu\nu} & = & r^\al_{\bt\mu\nu}(\Ga) =
\pt_\mu\Ga^\al_{\bt\nu}-\pt_\nu\Ga^\al_{\bt\mu} +
\Ga^\al_{\si\mu}\Ga^\si_{\bt\nu}-\Ga^\al_{\si\nu}\Ga^\si_{\bt\mu}
\nonumber\\
r_{\mu\nu} & = & r_{\mu\nu}(\Ga)=r^\al_{\mu\al\nu}
\eea
i.e. without assuming that $\Ga$ is the Levi-Civita
connection of $g$.

Unlike in a purely metric case, an equivalence with General Relativity  
also holds for non-linear gravitational Lagrangians
\be
L_f(g,\Ga)= \sqrt g\, f(r) 
\ee
(parameterized by the real function $f$ of one variable),
when they were considered within the first-order Palatini
formalism \cite{FFV}.
Similar analysis were performed for "Ricci squared"  non-linear 
Lagrangians
\be
\hat L_f(g,\Ga)=\sqrt g\, f(s)
\ee
where, $s=s(g,\Ga)=g^{\al\mu}g^{\bt\nu}r_{(\al\bt)}r_{(\mu\nu)}$, and
$r_{(\mu\nu)}=r_{(\mu\nu)}(\Ga)$ is the symmetric part of the Ricci 
tensor of $\Ga$. (Thereafter $()$ denotes a symmetryzation.)

Let us consider a $(1,1)$ tensor valued concomitant of a metric $g$ and
a linear torsionless connection $\Ga$ defined by
\be
S^\mu_\nu\equiv S^\mu_\nu (g, \Ga)=g^{\mu\la}r_{(\la\nu)} (\Ga)
\ee
One can use it to define a family of scalar concomitants of 
the Ricci type
\be
s_k=\tr S^k
\ee
for $k=1,\ldots n$.  
We can eliminate the higher order Ricci scalars $s_k$ with $k > n$,
by using a characteristic polynomial equation for the $n\times n$ 
matrix  $S$ (c.f. \cite{TU}). One immediately
recognizes that $r\equiv s_1=\tr S$ and $s\equiv s_2 =\tr S^2$.

\section{\hspace{-4mm}.\hspace{2mm}NONLINEAR RICCI LAGRANGIANS}

Our goal in the present note is to apply a Palatini principle to
the more general family of non-linear gravitational Lagrangians
of the Ricci type
\be
L_F(g,\Ga)=\sqrt g\, F(s_1,\ldots ,s_n)
\ee
parameterized by the real-valued function $F$ of $n$-variables. 
This family includes the previous ones as particular cases.

\subsection{\hspace{-5mm}.\hspace{2mm}Equation of Motion}
According to the Palatini prescription, we choose a metric $g$ and 
a symmetric connection $\Ga$ on a space-time manifold $M$ as independent 
dynamical variables. Variation of $L_F$ gives
\be
\de L_F  =  \sqrt g\,((\de_g F)_{\al\bt}-
\frac{1}{2}Fg_{\al\bt})\,\de g^{\al\bt} + \sqrt g\,\de_\Ga F
\ee
where  obviously $\de F=\sum_{k=1}^n F^\prime_k\,\de s_k$, and 
$F^\prime_k=\frac{\pt F}{\pt s_k}$.
We see at once that
$$
\de_g s_k= k\,\tr (S^{k-1}\de_g S)=
k\,(S^{k-1})^\si_\al\, r_{(\bt\si)}\,\de g^{\al\bt}
$$
which is clear from $\de s_k=k\,\tr (S^{k-1}\de S)$.
Accordingly
$$
\de_g F= ||F^\pr (S)||^\si_\al\, r_{(\bt\si)}\de g^{\al\bt}
$$
where for abbreviation we have introduced a $(1,1)$ tensor field concomitant
\be
||F^\prime (S)|| =\sum_{k=1}^n k\, F^\prime_k\, S^{k-1}
\ee
In a similar manner one calculates
\be
\de_\Ga F=||F^\pr (S)||^\al_\si\, g^{\si\bt}\,\de r_{(\al\bt)}
\equiv ||F^\pr (S)||^{\al\bt}\,\de r_{(\al\bt)}
\ee
where the inverse metric $g^{-1}$ has been used for rising the 
lower index in $||F^\pr (S)||$.
Substituting all necessary terms into formula (2.2) gives
\be
\de L_F =  \sqrt g\,(||F^\prime (S)||^\si_\al\, r_{(\bt\si)}-
\frac{1}{2}F g_{\al\bt})\,\de g^{\al\bt}- 
\sqrt g\, ||F^\prime (S)||^{\al\bt}\,\de r_{(\al\bt)}
\ee
Taking into account the well-known Palatini formula
$$
\de r_{(\al\bt)}=
\na_\mu\de\Ga^\mu_{\al\bt}-\na_{(\al}\de\Ga^\si_{\bt )\si}
$$
with $\na_\al$ being the covariant derivative with respect to 
$\Ga$ and performing the "covariant" Leibniz rule one gets the 
variational decomposition formula
\bea
\de L_F & = & \sqrt g\,(||F^\prime (S)||^\si_\al\, r_{(\bt\si)}-
\frac{1}{2}F g_{\al\bt})\,\de g^{\al\bt}- 
\na_\nu\, [\sqrt g\, (||F^\prime (S)||^{\al\bt}\,\de^\nu_\la 
\nonumber \\
&&{} - ||F^\pr (S)||^{\nu\al}\,\de^\bt_\la)]\,\de\Ga^\la_{\al\bt}+
\pt_\mu\, [\sqrt g\,||F^\prime (S)||^{\al\bt}\,
(\de\Ga^\mu_{\al\bt}-\de^\mu_{(\bt}\,\de\Ga^\si_{\al)\si})] 
\eea
This formula splits $\de L_F$
into the Euler-Lagrange part and the boundary term which shall be used
later on for a conserved current construction.

Therefore, the Euler-Lagrange field equations 
read as follows
\bea
||F^\prime (S)||^\si_{(\al}\,r_{(\bt)\si)}-\frac{1}{2}F\,g_{\al\bt}=0\\
\na_\nu\,[\sqrt g\,(||F^\pr (S)||^{(\al\bt)}\,\de^\nu_\la 
- ||F^\pr (S)||^{\nu (\al}\,\de^{\bt)}_\la)]=0
\eea

Before proceeding further, it is convenient to introduce a $(0,2)$
symmetric tensor field 
\be h_{\al\bt}=r_{(\al\bt)}(\Ga)\ee which will be extremely useful for 
studying symmetry properties of $||F^\pr (S)||$. For this purpose
we shall employ a matrix notation. For example: $S=g^{-1}\,h$ with 
both $g$ and $h$ being symmetric matrices (c.f. equation (1.4)), easily 
implies that $h\,S^k=g\,S^{k+1}$ and $S^k\,g^{-1}=S^{k+1}\,h^{-1}$ 
(provided that $h^{-1}$ exists) are also symmetric matrices for arbitrary 
$k=0,1,\ldots$. Indeed since e.g. $h\,S^k=h\,g^{-1}\ldots g^{-1}h$ then 
it is self-transpose. In particular, $h\,||F^\pr (S)||$ in (2.7) and
$||F^\pr (S)||\,g^{-1}$ in (2.8) (c.f. (2.4) and (2.11)) are symmetric. 
In other words e.g., the matrix concomitant
$$||F^\pr (S)||^{\al\bt}\equiv
%(||F^\pr (S)||\,g^{-1})^{\al\bt}\equiv 
||F^\pr (S)||^\al_\si\,g^{\si\bt}$$
is symmetric. These properties allow us to transform the Euler-Lagrange 
equations (2.7-2.8) into the form
\bea
S\,||F^\pr (S)|| =\frac{1}{2} F\, I\\
\na_\nu\,(\sqrt g\,||F^\prime (S)||^{\al\bt})=0
\eea
where $I$ is a $n\times n$ identity matrix. 
(Compare for similar calculations presented e.g. in [3-6,13,14].)

Equations (2.10) must be considered together with a consistency
condition obtained by taking the trace of (2.10). It gives
\be
\sum_{k=1}^n k\,F^\prime_k\, s_k=\frac{n}{2}\,F
\ee
The last equation (except the case it is identically satisfied)
becomes a single (non-algebraic in general) equation on possible 
values of the Ricci scalars (remember that $F$ and $F^\pr_k$ are given
functions of the variables $s_1,\ldots ,s_n$).
It forces $\,(s_1,\ldots ,s_n )\,$ to take a  set of constant values 
$s_i=c_i$, with $\,(c_1,\ldots ,c_n )\,$ being a solution of (2.12). 
Substituting back these constant roots into equation (2.10) we obtain
a polynomial equation for the matrix $S$. It means that with any
set $c_1,\ldots ,c_n$ of the (numerical) solutions  of (2.12),
one can associate a polynomial
\be
\pi_{c_1,\ldots , c_n} (t)=\sum_{k=1}^n a_k t^k
\ee
with  constant coefficients 
$a_k=k\,\frac{\pt F}{\pt s_k}\,(c_1,\ldots , c_n)$.
In other words, a lacking of an explicit dependence on a point $x\in M$
in equation (2.12), implies that the coefficients $a_i$ are also 
$x$-independent. The above arguments can be reinforce, following the 
line developed in \cite{TU}: by using the characteristic equation 
techniques, one is  allowed to introduce a complementary system
of $(n-1)$-equations that additionally relate  values of  the Ricci 
scalars and which still do not depend on a point $x\in M$. 
Thus, instead of the single equation
(2.12) we can have at our disposal a system of $n$-equations
with $n$-unknowns that provides us, in a regular case, in a set of numerical
( i.e. constant) solutions $(c_1,\ldots ,c_n)$.
But this rather technical point will be consider in more details 
elsewhere \cite{AB}. 

In this way we are led to the polynomial structure that has been 
defined at the very beginning. In our case the polynomial 
equation for $S$ takes the form
\be
S\, \pi_{c_1,\ldots , c_n} (S)= I
\ee
This becomes now a substitute of (2.10). (In fact, in order to get (2.14)
one eventually should rescale the coefficients in (2.13) by a 
constant factor.) Particularly, (2.14) implies that the determinant
of $S$ is a constant. As a consequence the determinant of 
$g$ is up to a constant factor proportional to that of $h$. 

From now on unless otherwise stated we assume that $S$ is an 
invertible matrix (nondegenerate case) with, of course, 
$S^{-1}= \pi_{c_1,\ldots , c_n} (S)$. Thus, replacing 
$\det\,g$ in (2.11) by $\det\,h$ and making use of the Ans\"{a}tz 
(2.9) with $h^{-1}=\pi (S)\,g^{-1}$ (c.f. (2.4)), gives
$$
\na_\la (\sqrt h h^{\al\bt})=0
$$
with $h^{\al\bt}$ being the inverse of $h_{\al\bt}$. This,
in turn, in any dimension $n>2$ 
\footnote{See \cite{FFV3,BFFV2} for $n=2$ case.},
forces $\Ga$ to be the Levi-Civita connection 
of $h$. Replacing back into (2.9) we find
\be
h_{\mu\nu}=r_{(\mu\nu)}(\Ga_{LC}(h))= R_{\mu\nu}(h)
\ee
the Einstein equations for the metric $h$. 
Here a value of the cosmological constant is $1$ due to the 
"unphysical" normalization made in (2.14). %constant $\La=1$
This shows that the use of Palatini formalism leads to results 
essentially different from the metric formulation when one deals 
with non-linear Ricci type Lagrangians: with the exception of special 
("non-generic") cases we always obtain the Einstein equations  as 
gravitational field equations. In this sense non-linear 
theories are equivalent to General Relativity (see also \cite{JK} 
in this context). They admit  alternative Lagrangians for 
the Einstein equations with a cosmological constant.

%\subsection{\hspace{-5mm}.\hspace{2mm}
%First Subsection of the Current
%Section (main words capitalized)}
\subsection{\hspace{-5mm}.\hspace{2mm}Symmetries and Superpotentials} 

Though the understanding of the energy of gravitational field has
not been attained yet, we can analyse the Noether symmetries and 
the corresponding conservation laws. 
Our Lagrangians are {\it reparameterization invariant}, 
in the sense that under a $1$-parameter group of diffeomorphisms 
generated by an arbitrary vector field $\xi=\xi^\al\pt_\al$ on $M$, 
the Lagrangian $L_F$ transform as a scalar density of weight $1$.  
At the infinitesimal level, variations of the field variables 
are represented by the Lie derivatives ${\cal L}_\xi$ , e.g.
$$
\de\Ga^\al_{\al\rho}\equiv {\cal L}_\xi \Ga^\bt_{\al\rho}
=\xi^\si R^\bt_{\al\si\rho} +\na_{\al}\na_{\rho}\xi^\bt\\
$$
(See also \cite{FF3} and \cite{BFF} for a self-contained exposition
of the Second Noether Theorem.)

The main contribution to the Noether current comes from
the boundary term in (2.6) that when expressed in terms of a new
metric (2.9) reads as follow
$$
\sqrt h\, h^{\al\bt}\,
(\de\Ga^\mu_{\al\bt}-\de^\mu_\bt\,\de\Ga^\si_{\al\si})] 
$$
As a consequence, one obtains the Komar expression 
\be
U_{F}^{\mu\nu}(\xi) = |\det h|^{\frac{1}{2}}
(\na^\mu\xi^\nu - \na^\nu\xi^\mu)
\ee
for a superpotential \cite{FF3,Kij,BFFV,Go,GS}
Therefore, an energy-momentum 
flow as well as a superpotential are already 
known from the standard Einstein-Hilbert formalism.
This extends a notion of universality for the Ricci 
type Lagrangians also to the energy-momentum complex \cite{BFFV2,BFFV}.

%\subsection{\hspace{-5mm}.\hspace{2mm}
%Related Differential - Geometric Structures}
\section{\hspace{-4mm}.\hspace{2mm}
RELATED DIFFERENTIAL - GEOMETRIC STRUCTURES}

The algebraic constraints (2.14) are of special interest by their own.
They provide on the space-time some additional 
differential-geometric structure, namely
a metric polynomial structure \cite{Opo}. A more complete treatment
of this subject will be done in a forthcoming publication \cite{AB}. 
For example, a polynomial structure related to the Lagrangians (1.2) is 
trivial and reduces into $S=I$. Therefore, both metrics $g$ and $h$ coincide
and we are left with purely Einstein equations. For the Lagrangians
(1.3), a polynomial structure turns out to be well-known
a pseudo Riemannian almost-product structure or/and an almost-complex 
anti-Hermitian ($\equiv$ Norden) structure  \cite{BFFV2}.  
Moreover, besides the initial metric $g$ one gets the Einstein 
metric $h$. Both metrics are related by algebraic equation  $S^2=\pm I$.
This was investigated in \cite{BFFV3}.

In the (psedo-)Riemannian almost-product case one equivalently deals 
with an almost-product structure given by the $(1,1)$ tensor field 
$S\equiv P$ ($P^2=I$) together with a compatible metric $h$
satisfying the condition
\be
h(PX, PY) = h(X, Y)
\ee
which is encoded in the simple algebraic relation (2.14).
(In our case the metric $h$ should be also Einsteinian.)
Here $X, Y$ denote two arbitrary vector fields on $M$.

There is a wide class of integrable
almost-product structures, namely so called {\it warped product} 
structures \cite{Besse,CC}, which are an intrinsic property of some 
well know exact solutions of Einstein equations: these include 
e.g. Schwarzschild, Robertson-Walker, Reissner-Nordstr\"{o}m, 
de Sitter, etc. (but not Kerr!).
Some other examples are provided by Kaluza-Klein type theories,
$3+1$ decompositions and more generally so called 
{\it split} structures \cite{GK}. The explicit form of the zeta
function on product spaces and of the multiplicative anomaly
has been derived recently in \cite{BW}.

In the anti-Hermitian case one deals with $2m$ - dimensional 
manifold $M$, an almost complex structure $S\equiv J$ ($J^2=-I$)
and an anti-Hermitian (Norden) metric $h$ \cite{GI}: \footnote{
Recall that for Hermitian metric $h(JX, JY) = h(X, Y)$.}
\be
h(JX, JY) = - h(X, Y)\ee
This implies that the signature of $h$ should be $(m,m)$.
In the K\"ahler-like case
($\na J = 0$ for the Levi-Civita connection of $h$) the almost-complex
structure is automatically integrable. We have proved  that in fact 
the metric $h$ has to  be a real part of certain  holomorphic 
metric on a complex (space-time) manifold $M$ \cite{BFFV3}. This leads
to a theory of anti-K\"{a}hler manifolds \cite{BFV}.

It should be also remarked that the theory of complex manifolds
with holomorphic metric (so called {\it complex Riemannian} manifolds) 
has become one of the corner-stone of the twistor theory \cite{Fl}.
This includes a {\it non-linear graviton} \cite{Pe}, 
{\it ambitwistor} formalism \cite{LeB}, theory of {\it $H$-spaces} \cite{BFP} 
or {\it Heavens} (i.e. self-dual holomorphic metrics) \cite{LeB}.

Of course, more general Ricci type Lagrangians (2.1) will produce, 
in general,  more complicated Einstein-metric-polynomial structures.
For example, the choice $F=s_3^2 \pm 16s_3$ in $n=4$ dimensions gives 
rise to the polynomial equation $S^3=\mp I$ \cite{AB}.

\end{document}